\documentclass[sigconf,screen,authorversion]{acmart}
\AtBeginDocument{%
  }

\setcopyright{cc}
\setcctype{by-nc-nd}
\acmDOI{10.1145/3832783.3834395}
\acmYear{2026}
\copyrightyear{2026}
\acmISBN{979-8-4007-2882-2/2026/10}
\acmConference[ASE '26]{Proceedings of the 41st IEEE/ACM International Conference on Automated Software Engineering}{October 12--16, 2026}{Munich, Germany}
\acmBooktitle{Proceedings of the 41st IEEE/ACM International Conference on Automated Software Engineering (ASE '26), October 12--16, 2026, Munich, Germany}
\acmSubmissionID{ase26main-p1571-p}
\received{2026-03-26}
\received[accepted]{2026-06-18}

\usepackage{xspace}
\usepackage{booktabs}
\usepackage{subcaption}
\usepackage{minted}
\usepackage{enumitem}
\setminted{
    fontsize=\scriptsize,
    escapeinside=||,
    breaklines,
    breaksymbolleft={},
    breaksymbolright={},
    breaksymbol={}
}
\usepackage{fontawesome5}
\usepackage{tabularx}
\usepackage{multirow}
\usepackage{multicol}
\usepackage{algorithm}
\usepackage{algpseudocode}
\usepackage[most]{tcolorbox}
\usepackage{wrapfig}

\AtBeginDocument{
  \setlength{\abovedisplayskip}{3pt}
  \setlength{\belowdisplayskip}{3pt}
  \setlength{\abovedisplayshortskip}{2pt}
  \setlength{\belowdisplayshortskip}{2pt}
}

\algblockdefx{Try}{EndTry}{\textbf{try}}{\textbf{end try}}
\algblockdefx{Catch}{EndCatch}[1]{\textbf{catch}~#1}{\textbf{end catch}}
\algrenewcommand\algorithmicrequire{\textbf{Input:}}
\algrenewcommand\algorithmicensure{\textbf{Output:}}
\algrenewcommand\algorithmicindent{1em}

\newcommand*\name{\textsc{ExplainBench}\xspace}
\newcommand*\auditagentname{\textsc{ExplanationAuditAgent}\xspace}

\newcommand{\ys}[1]{\textcolor{black}{#1}}

\begin{document}

\title{ExplainBench: Evaluating Code Explanations from Agents}

\author{Zhiyuan Pan}
\authornote{Work done while the author was in the National University of Singapore.}
\authornote{Three authors contributed equally to this research.}
\orcid{0009-0006-6059-5191}
\affiliation{%
  \institution{Zhejiang University}
  \city{Hangzhou}
  \country{China}
}
\email{zy\_pan@zju.edu.cn}

\author{Sungmin Kang}
\authornotemark[2]
\orcid{0000-0002-0298-5320}
\affiliation{%
  \institution{National University of Singapore}
  \country{Singapore}
}
\email{sungmin@nus.edu.sg}

\author{Imam Nur Bani Yusuf}
\authornotemark[2]
\orcid{0000-0003-0900-5230}
\affiliation{%
  \institution{National University of Singapore}
  \country{Singapore}
}
\email{inbyusuf@nus.edu.sg}

\author{Abhik Roychoudhury}
\orcid{0000-0002-7127-1137}
\affiliation{%
  \institution{National University of Singapore}
  \country{Singapore}
}
\email{abhik@nus.edu.sg}

\begin{abstract}
Large Language Model (LLM) agents have seen rapid adoption in software engineering. 
As agents take a greater role in the actual generation of code, they are making larger changes, spanning tens to hundreds of lines. This makes manual review of agent results increasingly infeasible, leading developers to turn to explanations to understand enacted changes. 
Despite this, there are no benchmarks that evaluate the trustworthiness of agent-generated explanations. To bridge this gap, we propose \name, a benchmark to automatically evaluate explanations from coding agents. \name is based on the intuition that informative explanations should enable an LLM to correctly answer questions, allowing quantitative comparison of explanation quality between agents.
With this observation, we construct a suite of questions that evaluates whether
explanations accurately describe (1) the intended behavior of buggy code and (2) the effect of applying the agent patch itself.
Experiments first reveal that explanation quality is a distinct axis of agent evaluation: \name ranks agents differently from the widely-used SWE-bench Verified benchmark. A deeper breakdown of explanation quality in agents shows frequent problems in explanations, such that explanations often claim that a patch is correct when it is not. Based on this insight, we implement and evaluate an explanation audit agent which runs additional tests to validate and refine explanations. This agent improved the explanations of all evaluated agents, demonstrating agent explanations can be automatically made more trustworthy.
\end{abstract}




\maketitle

\section{Introduction}
\label{sec:introduction}

Large Language Model (LLM) agents are LLMs equipped with a scaffolding that allows them to invoke tools to retrieve information and enact changes. Agents provide state-of-the-art performance in software engineering tasks such as test generation~\cite{ahmed2025otter} or program repair~\cite{yang2024sweagent}, and are reshaping the practice of software engineering. This has led to their quick adoption by industry and a change in how developers work. For example, 
Anthropic reported that a majority of employees were using Claude Code on a daily basis~\cite{huang2025aiwork}.

As LLM agents become more flexible and deal with complex tasks, it becomes increasingly important that agents provide \emph{trustworthy explanations}. This is because the speed and scale of agent code generation make manual inspection of agent results increasingly costly, while explanations provide a useful summary of the implementation, provided the explanation is reliable. Thus, as the capability of LLM agents grows, developers would increasingly treat agents like junior developers summarizing their work -- making it important for agents to provide accurate and honest explanations, in addition to being capable in software engineering tasks.

However, the importance of explanations for agents has not yet been recognized, and there is a lack of benchmarks that evaluate explanation trustworthiness. This is in contrast to the large number of benchmarks evaluating agent efficacy~\cite{jimenez2024swebench, mundler2024swtbench, jing2024dsbench}. This absence poses two problems. First, it is difficult to tell which agent explanations are the most reliable, as opposed to which agents are the most effective. Secondly, the lack of a benchmark makes it difficult for researchers to iterate on their design and find techniques that can improve the correctness of patch explanations.

This work introduces \name, a benchmark that automatically and quantitatively assesses the quality of patch explanations generated by agents. The key challenge in evaluating explanation quality is that agent explanations are in natural language, making automated and objective assessment of the content difficult. To overcome this challenge, \name evaluates the natural language explanation in the form of an \emph{LLM questionnaire}. In particular, \name provides an explanation as input to an LLM, and has the LLM answer multiple-choice questions about the bug and patch. The intuition is that informative explanations will be sufficient to correctly answer questions, while vacuous explanations with no meaningful content will cause an LLM to answer questions incorrectly. Quantitatively, we use the proportion of correctly answered questions as the explanation score. In turn, this score allows comparison between agent explanations, while facilitating experimentation to improve agent explanations. %

Evaluating five agents on \name first revealed that agent explanation quality is independent from agent efficacy. For example, the trae-agent~\cite{traeresearchteam2025traeagentllmbasedagent} in our evaluation had the highest SWE-bench Verified resolution rate, but had the second-lowest explanation score on \name. 
Further analysis reveals that existing agent explanations are over-optimistic about the correctness of the patch, and suffer from inaccurate inference of function-level intended behavior.
To address these issues, we propose \auditagentname, an LLM agent that runs differential testing \ys{on the code before and after the patch} and audits claims in agent explanations. The audit agent could generate higher-quality refined explanations for all agents we experiment with, demonstrating its capability as a general explanation improver for any software agent.

Our contributions are as follows:
\begin{itemize}[leftmargin=*]
    \item We propose \name, which automatically evaluates agent explanation quality.
    \item We find patch efficacy does not always correspond with explanation quality, showing the need for evaluation of explanations.
    \item We implement \auditagentname, which performs differential testing based on the content of the explanation and provides a refined and more trustworthy explanation.
\end{itemize}

While we evaluate certain well-defined aspects of explanations, we do not argue that these are the only quality aspects that can be evaluated. In particular, we clarify that \name evaluates explanation correctness in terms of explanation content, which is a key issue in software documentation~\cite{aghajani2019software}; it does not measure developer satisfaction or readability. We publish our code~\cite{explainbench_package}, written in a modular fashion, such that the benchmark can be extended to support the evaluation of explanations on criteria other than those used in this paper.

\section{Related Work}
\label{sec:relwork}

\subsection{Agents for Software Engineering}

LLMs have been rapidly adopted in many industries, but their application in software engineering has been notably widespread. Among the first agents for software engineering were the issue-resolving agents SWE-agent~\cite{yang2024sweagent} and AutoCodeRover~\cite{Zhang2024autocoderover}, which provided a scaffolding around LLMs to autonomously retrieve information and generate patches. Subsequently, other domain-specific agents have been proposed, such as the Otter~\cite{ahmed2025otter} agent for generating bug-reproducing tests or the ExecutionAgent~\cite{Bouzenia2025ExecutionAgent} agent for building execution environments. In the opposite direction, more general agents which can handle many types of tasks such as OpenHands~\cite{wang2024openhands} or USEAgent~\cite{applis2025useagent} have also been explored. Despite such domain and architecture diversity, and the frequency and ease with which they provide natural language explanations, the trustworthiness of their explanations is poorly understood.

\subsection{Benchmarks for Agentic SE}

Many benchmarks evaluating agents on software engineering tasks have been proposed. A prominent benchmark is SWE-bench Verified~\cite{chowdhury2024swebenchverified}, which consists of 500 human-verified instances from the earlier SWE-bench~\cite{jimenez2024swebench} benchmark of reproducible real-world issues. Influenced by SWE-bench Verified, many other benchmarks that evaluate agent code generation capability have been proposed. SWE-Poly Bench~\cite{rashid2025swepolybench} and Multi-SWE-bench~\cite{zan2025multiswebench} are multilingual variants on SWE-bench. SWE-bench-Live~\cite{zhang2025swebenchlive} and SWE-Bench-CL~\cite{joshi2025swebenchcl} order bugs by time: SWE-bench-Live updates regularly to avoid LLM contamination, while SWE-Bench-CL evaluates the ability of agents to utilize prior issues. Benchmarks in specialized domains have also been proposed. DSBench evaluates the ability of agents to perform data science tasks~\cite{jing2024dsbench}, while Terminal-Bench evaluates the ability of agents to perform complicated tasks in the terminal~\cite{tbench_2025}; GitTaskBench~\cite{ni2025gittaskbenchbenchmarkcodeagents} proposes a collection of realistic tasks across different modalities and domains.

To the best of our knowledge, \name is the first benchmark to automatically evaluate agent explanation quality. In this work, we compare explanation quality with issue resolution rate on SWE-bench Verified to show that explanation quality is an independent axis of evaluation.

\subsection{Explanations in Software Engineering}

Surveys on developers have revealed a consistent demand for explanations when using automated software engineering techniques. Noller et al.~\cite{noller2022trust}, who surveyed developer perception of what is needed to trust program repair tools, found that explanations were the second most sought after artifact (after patches themselves) for program repair tools. Developers also indicate that explanations would be used extensively in their reasoning process: one developer survey by Kochhar et al.~\cite{Kochhar2016FLSurvey} noted that they would judge the result in the context of the explanation.

While earlier surveys on automated debugging techniques noted an absence of explainable techniques, the introduction of LLMs has significantly changed the situation. LLMs, which are capable of generating both source code and natural language, have been frequently deployed to make claims to explainability. As a result, there are several works~\cite{kang2025autosd, Widyasari2024ExplainableFL, Mao2025ExplainableVulnerability} that uses LLMs to generate explanations along with patches and fault localization results. However, due to an absence of a framework to evaluate explanations, each research paper had to perform an ad-hoc manual evaluation of explanations. While this is qualitatively revealing, the lack of quantitative evaluation makes it difficult to compare explanation generation techniques. Moreover, as a natural byproduct of their chain-of-thought operation, many agents also generate natural language explanations, as described in Section~\ref{sec:pipeline:expl_extraction}, which have not yet been evaluated to the best of our knowledge.

\begin{figure}[t!]
    \centering
    \begin{subfigure}[b]{0.49\textwidth}
        \centering
        \begin{minted}[fontsize=\scriptsize]{markdown}
### Issue Requirements Check
**Original Issue**: Setting `min-similarity-lines` to
0 should disable the duplicate code check  
...
### Final Verification
The fix successfully addresses the GitHub issue by:
1. **Disabling the duplicate code check** when `min-similarity-lines=0`
...
The fix resolves the issue completely (...)
        \end{minted}
        \caption{Natural language explanation.\label{fig:motivating_example_1__explanation}}
    \end{subfigure}
    \begin{subfigure}[b]{0.49\textwidth}
        \centering
        \begin{minted}[fontsize=\scriptsize]{diff}
diff --git a/pylint/checkers/similar.py b/pylint/checkers/similar.py
--- a/pylint/checkers/similar.py
+++ b/pylint/checkers/similar.py
@@ -830,6 +830,8 @@ class SimilarChecker(BaseChecker, Similar, MapReduceMixin):
...
+ if self.min_lines <= 0:
+   return
...         
        \end{minted}
        \caption{Submitted patch (never covered).\label{fig:motivating_example_1__patch}}
    \end{subfigure}
    \caption{Example explanation and submitted patch from the Lingxi agent~\cite{yang2025lingxirepositorylevelissueresolution}.}
    \label{fig:motivating_example_1}
    \vspace{-0.3cm}
\end{figure}

\section{Motivation}
\label{sec:motivating_example}

To illustrate why agent explanations need to be evaluated, we discuss the motivating example in Fig.~\ref{fig:motivating_example_1}, based on output from the Lingxi agent~\cite{yang2025lingxirepositorylevelissueresolution}. When interacting with agentic systems, developers often see a natural language summary before inspecting the generated patch. For instance, the popular Claude Code\footnote{\url{https://claude.com/product/claude-code}} tool presents a natural language summary upon completion. In our example presented at Fig.~\ref{fig:motivating_example_1__explanation} as well, a developer would read the summary and naturally assume that a patch was successfully implemented and resolved the issue. Even if they saw the patch (Fig.~\ref{fig:motivating_example_1__patch}), it would appear reasonable. However, a developer in this example would be surprised: the patch-modified method is actually never covered in bug-reproducing tests, indicating that the patch has no impact on the bug. Whenever developers experience such an explanation-behavior mismatch, their \emph{trust} in agentic systems erodes.

\begin{figure*}
    \centering
    \includegraphics[width=0.95\linewidth]{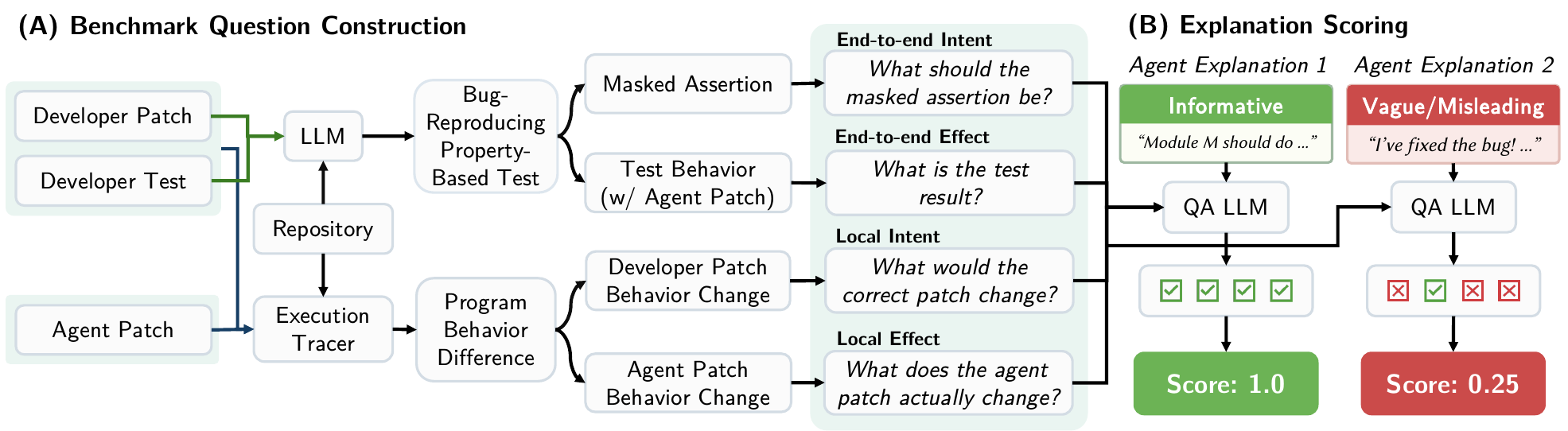}
    \caption{Overview of \name.}
    \label{fig:overall_diagram}
    \vspace{-0.3cm}
\end{figure*}

To assess the extent of such problems, this work builds a benchmark to evaluate to what degree agents would generate inaccurate explanations. 
The main challenge in automatically evaluating explanations is that the explanations provide information in natural language, and thus the same information may take many different forms, making it difficult to automatically evaluate explanations. To overcome this challenge, we formulate a \emph{questionnaire} with verifiable answers, and check if the natural language explanation provides sufficient information for an LLM to correctly answer questions. See Fig.~\ref{fig:overall_diagram}(B) for a schematic of our core intuition: informative explanations would allow an LLM to correctly answer questions about the bug and patch, while uninformative or misleading explanations would lead to incorrect answers. In turn, based on the accuracy of the questionnaire, one can deduce how informative an explanation is. Through careful choice of the question types, one can evaluate whether specific types of information are present in the explanation as well, allowing a fine-grained assessment of explanation quality.

In summary, our \emph{problem statement} is that we are interested in automatically evaluating the quality of explanations generated by coding agents, making quality comparison and improvement feasible. Our \emph{core assumption} to solve this problem is that informative explanations would help LLMs accurately answer questions about the bug. LLMs are an imperfect yet useful evaluator of natural language artifacts, as evidenced by the large number of benchmarks using them~\cite{li2024crowdsourced,arenahard2024, DBLP:conf/nips/ZhengC00WZL0LXZ23}; LLMs are particularly adept at identifying relevant information from text~\cite{DBLP:journals/corr/abs-2403-05530}, relevant to our setup. In this work, we have the LLM answer questions with objective answers as an additional measure of rigor.

\section{\name Framework}
\label{sec:framework}

This section describes the general structure of \name and what questions it generates; the detailed question construction process is described in Section~\ref{sec:pipeline}.

\subsection{Overview}

Fig.~\ref{fig:overall_diagram} presents an overview of our evaluation framework, \name. Our goal is to assess whether a patch explanation conveys information relevant to (1) the intended change in program behavior and (2) the actual effect of the agent patch. To this end, \name first gathers behavioral information from developer tests, developer patches and agent patches. \name then constructs questions about the bug and patch (Fig.~\ref{fig:overall_diagram}(A)). Finally, explanations are evaluated by providing them to an LLM, and having the LLM answer the questions (Fig.~\ref{fig:overall_diagram}(B)).

The remainder of this section describes the principles behind the benchmark construction and the question types used in our study. The specific process of artifact preparation and question construction on top of the SWE-bench Verified~\cite{jimenez2024swebench} benchmark is described in Section~\ref{sec:pipeline}.

\subsection{Principles}
\label{sec:methodology:setup}

\name evaluates explanations by running questions against the explanations using an LLM. In constructing these questions for \name, we followed the following principles:

\begin{enumerate}[leftmargin=*]
    \item \textbf{When possible, focus on questions answerable with general expressions, not specific values.} Questions with values as answers, as opposed to symbolic \emph{expressions}, require explanations to contain unnecessary details about how to calculate the expected value. In contrast, explanations should concisely describe key characteristics of the bug, which can be translated by an LLM into expressions.
    \item \textbf{LLMs should not directly synthesize expressions when answering questions.} Generating valid code expressions requires a significant amount of context. It is thus unreasonable to expect that an explanation should contain all of this only to reconstruct expressions.
\end{enumerate}

These principles motivate our use of \emph{multiple-choice questions (MCQs)} in \name. \name automatically generates validated expressions that capture the essence of both the developer patch and agent patch. This avoids the direct use of complex values and the synthesis of expressions during question-answering.

\subsection{Evaluation Desiderata}

This work evaluates whether explanations contain the \emph{intent} and \emph{effect} of the agent patch. Intent refers to what the agent patch should do, while effect refers to what the agent patch actually does. While intent and effect would ideally be identical, they can be different in practice; good explanations would distinguish the two. In our setting, questions about intent are derived from the developer patch, which we consider an oracle for the developer intent, as is done in prior work~\cite{wang2020automated}. Meanwhile, questions about effect are derived from the agent patch, representing the actual change to the code.

We focus on intent and effect in this work based on prior research on how developers document changes. First, intent is often missing yet desired from LLM-generated explanations: Xiao et al.~\cite{xiao2024generative} find that one of the common manual edits to LLM-generated pull request descriptions was to add the intent of the change. Prior work on commit messages and pull requests~\cite{tian2022makes, li2022follow} also notes the need to describe the expected behavior of the code. Meanwhile, the explanation needs to accurately describe the actual effect of a patch to help developers understand whether to accept patches~\cite{liang2019explain}. 

Meanwhile, we further divide evaluation of intent and effect by granularity, namely based on an end-to-end scope and a local scope. Ideally, an explanation would describe how the program works as a whole. Furthermore, developers often need to know which program component is behaving incorrectly and what its intended behavior should be~\cite{Kochhar2016FLSurvey}. While there are different granularities one could use for local questions, we opt to focus on the function level. This is because existing studies suggest that functions are cohesive logical units of software~\cite{braver2022functions}, making them the most useful unit to explain.

\begin{figure}
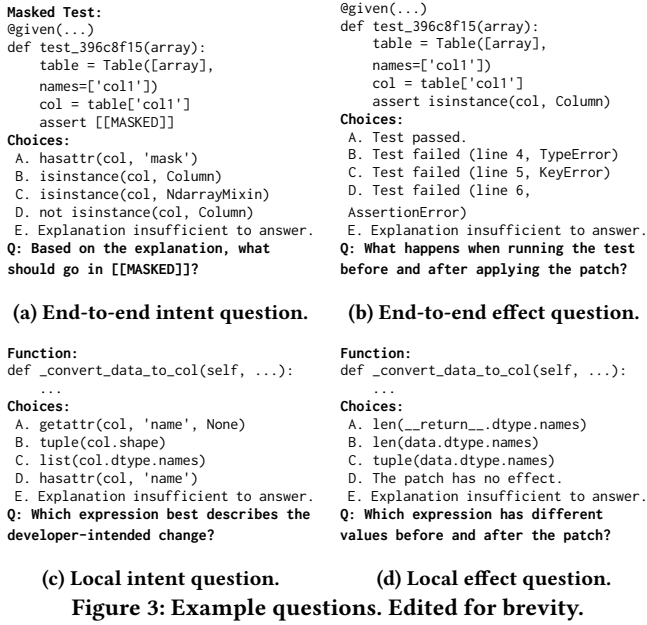

    \centering
    \begin{subfigure}[b]{0.48\columnwidth}
        \centering
        \begin{minted}[escapeinside=||]{text}
|\textbf{Masked Test:}|
@given(...)
def test_396c8f15(array):
    table = Table([array], names=['col1'])
    col = table['col1']
    assert [[MASKED]]
|\textbf{Choices:}|
 A. hasattr(col, 'mask')
 B. isinstance(col, Column)
 C. isinstance(col, NdarrayMixin)
 D. not isinstance(col, Column)
 E. Explanation insufficient to answer.
|\textbf{Q: Based on the explanation, what should go in [[MASKED]]?}|
        \end{minted}
        \caption{End-to-end intent question.}
        \label{fig:example_qs__e2eintent}
    \end{subfigure}
    \hfill
    \begin{subfigure}[b]{0.48\columnwidth}
        \centering
        \begin{minted}{text}
|\textbf{Test:}|
@given(...)
def test_396c8f15(array):
    table = Table([array], names=['col1'])
    col = table['col1']
    assert isinstance(col, Column)
|\textbf{Choices:}|
 A. Test passed.
 B. Test failed (line 4, TypeError)
 C. Test failed (line 5, KeyError)
 D. Test failed (line 6, AssertionError)
 E. Explanation insufficient to answer.
|\textbf{Q: What happens when running the test before and after applying the patch?}|
        \end{minted}
        \caption{End-to-end effect question.}
        \label{fig:example_qs__e2eeffect}
    \end{subfigure}

    \vspace{0.2cm}

    \begin{subfigure}[b]{0.48\columnwidth}
        \centering
        \begin{minted}{text}
|\textbf{Function:}|
def _convert_data_to_col(self, ...):
    ...
|\textbf{Choices:}|
 A. getattr(col, 'name', None)
 B. tuple(col.shape)
 C. list(col.dtype.names)
 D. hasattr(col, 'name')
 E. Explanation insufficient to answer.
|\textbf{Q: Which expression best describes the developer-intended change?}|
        \end{minted}
        \caption{Local intent question.}
        \label{fig:example_qs__localintent}
    \end{subfigure}
    \hfill
    \begin{subfigure}[b]{0.48\columnwidth}
        \centering
        \begin{minted}{text}
|\textbf{Function:}|
def _convert_data_to_col(self, ...):
    ...
|\textbf{Choices:}|
 A. len(__return__.dtype.names)
 B. len(data.dtype.names)
 C. tuple(data.dtype.names)
 D. The patch has no effect.
 E. Explanation insufficient to answer.
|\textbf{Q: Which expression has different values before and after the patch?}|
        \end{minted}
        \caption{Local effect question.}
        \label{fig:example_qs__localeffect}
    \end{subfigure}

    \vspace{-0.4cm}
    \caption{Example questions. Edited for brevity.}
    \label{fig:example_qs}
    \vspace{-0.4cm}
\end{figure}

\subsection{Question Types}

\label{sec:methodology:formats}
Based on evaluation desiderata, \name contains four questions from the intersection of intent/effect and end-to-end/local, as described next. Examples for each question are given in Fig.~\ref{fig:example_qs}.

\subsubsection{End-to-End Intent}
This question type evaluates whether an explanation conveys what the program is intended to do in contrast to the buggy behavior. This may be understood as the program-level specification that the buggy implementation violates.

\noindent\textbf{\textit{Question Format.}} An LLM is given an agent explanation as well as a \emph{masked} bug-reproducing property-based test (PBT). Property-based tests are tests that sample inputs satisfying a precondition, then check that program output follows a property. This makes PBTs well-suited to our design principles~\cite{claessen2000quickcheck}. In the question, the PBT is masked so that the expression describing the expected behavior must be guessed. The LLM is then asked a multiple-choice question to determine which candidate expression would make the PBT fail on the buggy version and pass on the fixed version. This question thus evaluates whether the explanation provides sufficient information for an LLM to identify the expression that correctly describes the intended behavior change of the program as a whole.

\subsubsection{End-to-End Effect}

This question type evaluates whether an explanation sufficiently describes how the agent-submitted patch changes the behavior of the program. The reliability of this information has important implications for real-world use: if an agent reports that it has `fixed a bug', how much should this be trusted?

\noindent\textbf{\textit{Question Format.}} An LLM is given an agent explanation as well as a bug-reproducing PBT. The LLM is then asked a multiple-choice question on the result of running the test (i) before the patch and (ii) after the patch is applied. The options given are: the test passed, a few options stating the test failed at line \texttt{N} with exception \texttt{Exception}, or that there is insufficient information in the explanation to answer the question. This question thus evaluates whether an explanation is enough to derive how the program behavior would change, as revealed by a test.

\vspace{-0.2cm}
\subsubsection{Local Intent}
This question type evaluates whether an explanation describes the expected behavior of the faulty component at the function level. This may be understood as evaluating whether the explanation correctly pinpoints the faulty component and describes what function-level specification was violated.

\noindent\textbf{\textit{Question Format.}}
An LLM is given an agent explanation as well as (1) a function influenced by the bug (selection process described in Section~\ref{sec:pipeline}), (2) a source line where the bug manifests, and (3) the inputs to the function.
For each local question, the provided context also specifies whether expressions are evaluated immediately before or after executing the specified line; Fig.~\ref{fig:example_qs} omits the line and timing context for space.
The LLM is then asked the question \emph{``Within the context of the provided function and inputs, at the specified timing point relative to the specified line, which expression best describes the developer-intended change?''}

\vspace{-0.2cm}
\subsubsection{Local Effect}
\label{sec:question_type__local_effect}
This question type evaluates whether the explanation describes the actual effect of the agent patch at the function level. 
This information is important to let the end-users know how the patch influences the program state at the function level.

\noindent\textbf{\textit{Question Format.}}
An LLM is given an agent explanation as well as (1) the pre-patch version of a function influenced by the bug (selection process described in Section~\ref{sec:pipeline}), (2) a source line where the bug manifests, and (3) the inputs to the function.
We do not provide the agent-generated patch, which prevents the question-answering LLM from inferring local changes directly from the patch rather than from the explanation.
The LLM is then asked: \emph{``Within the context of the provided function and inputs, at the specified timing point relative to the specified line, which expression has different values before and after the patch?''}

\subsection{Evaluation Metrics}
During explanation evaluation, each patch explanation is paired with each of the questions, and an LLM is prompted to answer the question based on the explanation. As the questions are in MCQ format, the LLM generates a single-letter answer (or a pair of letters for end-to-end effect questions). The question is considered correctly answered when the answer from the LLM is identical to the letter(s) corresponding to the ground-truth.

To quantitatively evaluate the explanation quality of a coding agent, we define the \emph{explanation score} based on whether the explanation is sufficient to correctly answer the questions. Simply put, the explanation score is the proportion of questions from \name that were correctly answered. Formally, let $d \in \mathcal{D}$ denote an evaluation dimension (question type) and let $a_{b,d} \in \{0,1\}$ indicate whether the explanation for bug instance $b \in \mathcal{B}$ leads to an LLM correctly answering the question associated with dimension $d$. The explanation score is computed using Equation~\ref{eq:explanation_score}. 
\begin{gather}
\label{eq:explanation_score}
S_{\mathrm{exp}} =
\frac{1}{|\mathcal{D}|}
\sum_{d \in \mathcal{D}}
\left(
    \frac{1}{|\mathcal{B}|}
    \sum_{b \in \mathcal{B}} a_{b,d}
\right)
\end{gather}
While Equation~\ref{eq:explanation_score} averages over all question types, one can get scores for individual question types as well, facilitating deeper analysis of the explanations.

\section{\name Construction}
\label{sec:pipeline}

We next describe how the four types of questions introduced in the previous section are implemented in \name. We build our benchmark on top of SWE-bench Verified~\cite{jimenez2024swebench}, a benchmark of human-validated open-source issue descriptions and patches. Given a bug instance from SWE-bench Verified and an extracted agent explanation, our construction pipeline derives multiple-choice questions with correct answers that are grounded either in developer patches (for intent questions) or agent patches (for effect questions).

\subsection{Agent Explanation Extraction}
\label{sec:pipeline:expl_extraction}

First, we establish that modern LLM agents often include a natural language explanation along with their final answer in the last tool call step. Specifically, we find that 9 of the top 10 agents on the SWE-bench Verified leaderboard\footnote{URL: \url{https://www.swebench.com}, February 2026.} generate natural language explanations in the last step, based on public submissions\footnote{\url{https://github.com/SWE-bench/experiments}}.

To evaluate these explanations with \name, we first extract the natural language explanations produced by coding agents during bug-fixing. This is done by parsing agent trajectories to locate the final tool-call step, then collecting the associated natural-language rationale without modification. This ensures that our evaluation is performed directly on the explanations generated by the agents themselves.%

\subsection{End-to-End Question Construction}

\subsubsection{Property-based Test Construction}

As previously described, end-to-end questions use bug-reproducing property-based tests (PBTs) as a symbolic representation of the input-output behavior of the program. As the SWE-bench Verified benchmark does not provide bug-reproducing PBTs, we needed to generate such PBTs. To do so, we used the test generation agent of SpecRover~\cite{Ruan2025SpecRover}, which takes an issue statement and writes a bug-reproducing test, but with two changes. First, the agent additionally took the developer patch and (example-based) bug-reproducing test as input to aid PBT generation. Second, the agent was prompted to provide a PBT as output, instead of an example-based test. All resulting tests were validated such that they fail on the buggy version and pass on the developer-fixed version. While this process could automatically generate bug-reproducing PBTs for 98\% of the bugs in \name, in 2\% of cases, PBT generation failed despite repeated attempts. In such cases, we manually wrote PBTs as necessary. %

\subsubsection{End-to-End Intent}

Explanations should contain sufficient information to reconstruct the expression representing the expected behavior among competing candidates. Thus, we first manually identified the expected behavior expression from the generated PBTs for each bug. In the question, this expression is masked, so that the LLM must choose from a set of expressions to complete a bug-reproducing test. This leaves us with a masked PBT and correct expected behavior expression. However, as our questions are in MCQ format, incorrect answers must also be generated. This is done in two steps. First, an LLM is prompted to generate `distractors', i.e., incorrect expected behavior expressions similar to the correct expression. Next, these `distractor' expressions are validated to confirm that they are genuinely not bug-reproducing. Specifically, the distractor expressions are replaced with the correct expression in the PBT, and the PBT is executed. Distractor expressions that make the PBT pass on the buggy version or fail on the fixed version are kept as validated distractors. With a set of validated distractors, the question is composed with the following components: (1) a masked PBT with the correct expression replaced by \texttt{[[MASKED]]}, (2) choices, including the correct answer, the distractors, and one `explanation insufficient to answer' option for analysis, and (3) the question, asking the LLM what expression should go in \texttt{[[MASKED]]}. An example of the end-to-end intent question is shown in Fig.~\ref{fig:example_qs__e2eintent}.

\subsubsection{End-to-End Effect}

While the intent of the code stays constant, as it is defined by the developer patch, the effect of each agent patch differs by agent. As such, end-to-end effect questions need to be generated for each agent submission.

To construct the questions, we first record whether the PBT passes or fails before and after applying the patch. When the test fails, information on which line causes the failure and what exception was triggered is recorded. Note that the behavior may be equivalent before and after the patch, if the agent patch is disconnected with the bug. In addition to behavior from the buggy and patched code, $N_{\text{distractor}}$ distractor options are generated by prompting an LLM to generate alternative lines and exceptions at which the test could fail. With a set of distractors, the question is composed with the following components: (1) the full PBT, (2) choices, including the true pre-patch and post-patch behavior, distractors, and one `explanation insufficient to answer' option for analysis, and (3) the question asking the test execution result before and after the patch. An example question is shown in Fig.~\ref{fig:example_qs__e2eeffect}.

\subsection{Local Question Construction}

For local questions, distinguishing the execution behavior of the buggy and patched programs under the same test input is crucial.
Behavioral differences can be represented via expressions that change in value before and after a patch (or the absence of such expressions for ineffective patches). 
We formalize this as finding \textit{delta behavior} from the execution of the buggy and patched programs. Specifically, given a test input $t$, let $T_B = \langle (s_0, \sigma_B(0)), (s_1, \sigma_B(1)), \ldots \rangle$ and $T_P = \langle (s_0, \sigma_P(0)), (s_1, \sigma_P(1)), \ldots \rangle$
denote the execution traces of a buggy program $B$ and a patched program $P$, where $s_{\ell}$ is the $\ell$-th executed statement and $\sigma_{X}(\ell)$ is the variable state after executing $s_\ell$ in program $X \in \{B, P\}$. The delta behavior is defined as the smallest index $\ell$ where the executions differ in either variable state ($\sigma_B(\ell) \neq \sigma_P(\ell)$) or control-flow ($s_\ell(B) \neq s_\ell(P)$). We choose the earliest index $\ell$ because this location represents the beginning of all subsequent behavioral differences. 

Given the identified delta behavior, we generate expressions characterizing the behavioral difference as question choices, based on our principles in Section~\ref{sec:methodology:setup}. 
The ground-truth value of each expression is evaluated at the specified timing point, immediately before or after executing the line associated with the delta behavior.
Examples of local intent and effect questions are shown in Figs.~\ref{fig:example_qs__localintent} and~\ref{fig:example_qs__localeffect}. The choices in the questions consist of: (1) the correct answer, (2) the distractors, and (3) one ``\textit{explanation insufficient to answer}'' option for analysis. For local effect questions, we additionally include a ``\textit{patch has no effect}'' option. 
The key difference between intent and effect questions is that
intent questions are constructed based on the delta behavior from developer patches, whereas effect questions are constructed based on the delta behavior from agent patches.

\subsubsection{Execution Trace Collection}

To compare the execution behavior of buggy and patched programs, it is essential to record both control and data flow. Therefore, we collect execution traces by implementing a Python execution tracer based on the \texttt{sys.settrace()} function, as is common in prior benchmarking work~\cite{chen2025reasoning}.
Specifically, we instrument the official SWE-bench test execution harness\footnote{\url{https://github.com/SWE-bench/SWE-bench}}.
As the instrumented pipeline executes the test suites on both program versions, the tracer records execution events  (\textit{control flow}) and variable values (\textit{data flow}) at line-level granularity.
To difference variables in Python, we implement a serializer embedded in the tracer that converts objects into JSON representations. To reduce the overhead of instrumentation, we prescreen each test case to identify all 
callers of patch-modified functions, and restrict heavyweight tracing to these filtered functions.
The recorded events are subsequently used to construct a hierarchical representation of function frames, forming the basis for comparing buggy and patched program executions and for identifying the behavioral deltas between their corresponding traces in the next step.

\subsubsection{Delta Behavior Extraction}

With execution traces collected in the previous step, we extract \emph{delta behavior}. As defined above, delta behavior is the first semantically observable difference between pre- and post-patch executions, manifesting either as a difference in variable state at the $\ell$-th executed statement ($\sigma_B(\ell) \neq \sigma_P(\ell)$) or as a discrepancy in control-flow ($s_\ell(B) \neq s_\ell(P)$).

\begin{algorithm}[t]
\caption{Delta Behavior Extraction}
\label{alg:delta}
\footnotesize
\setlength{\abovedisplayskip}{2pt}
\setlength{\belowdisplayskip}{2pt}
\begin{algorithmic}[1]
\Require Buggy trace $T_B$, Patched trace $T_P$
\Ensure Delta behavior $\Delta$ or $\emptyset$
\Statex \hspace*{-1.65\algorithmicindent}\textbf{Notation:} $f$: \emph{function}

\State $f_B \gets$ entry frame of $T_B$; $f_P \gets$ entry frame of $T_P$

\While{$f_B \neq \emptyset$ \textbf{and} $f_P \neq \emptyset$}

    \If{$\mathit{hasNext}(f_B)$ \textbf{and} $\mathit{hasNext}(f_P)$}
        \State $s_B \gets \mathit{next}(f_B)$; $s_P \gets \mathit{next}(f_P)$ 
    \Else
        \State $f_B \gets f_B.\mathit{caller}$; $f_P \gets f_P.\mathit{caller}$
        \State \textbf{continue}
    \EndIf

    \If{$s_B = s_P$} {\color{gray}\Comment{control-flow match case}}
        \If{$s_B$ is a function call}
            \State $f_B \gets f_B.\mathit{step\_into}(s_B)$; $f_P \gets f_P.\mathit{step\_into}(s_P)$
        \ElsIf{$\sigma_B \neq \sigma_P$} {\color{gray} \Comment{variable state differs}}
            \State \Return observed variable difference
        \EndIf
    \Else {\color{gray} \Comment{control-flow mismatch case, $s_B$ $\neq$ $s_P$}}
        \If{outcomes of $f_B$ and $f_P$ differ}
            \State \Return observed outcome difference
        \EndIf
        \State $f_B \gets f_B.\mathit{caller}$; $f_P \gets f_P.\mathit{caller}$
    \EndIf

\EndWhile
\State \Return $\emptyset$
\end{algorithmic}
\end{algorithm}

Algorithm~\ref{alg:delta} summarizes our procedure for extracting delta behaviors from a pair of buggy and patched execution traces. The algorithm takes as input the buggy and patched program execution traces, $T_b$ and $T_p$ respectively, and returns either the first semantically observable behavioral difference or $\emptyset$ if no difference is observed.

The algorithm traverses both the buggy and patched execution traces from their entry frames while preserving calling context.
At each step, it advances by retrieving aligned execution events when available, steps into callee frames for function-call events, and compares observable program states for non-call events, reporting the first detected difference as the delta behavior. Misaligned frames or events are treated as control-flow divergences, in which case the algorithm compares function outcomes (return values or exceptions) and either reports a discrepancy or backtracks to caller frames to continue the traversal. If the traversal completes without detecting any difference, the algorithm returns $\emptyset$, indicating equivalent observable behavior under the given test input.

\subsubsection{Candidate Expression Generation and Validation}

After the delta behavior is extracted, the next step is to generate candidate expressions that capture the behavioral change the patch causes, along with unchanged expressions (distractors).%
Given the delta behavior, we prompt a separate LLM to generate candidate changed and unchanged expressions based on the following information: (1) the function in buggy and patched versions in which the divergence occurs, (2) the diverging statement identified by Algorithm~\ref{alg:delta}, and (3) the complete variable states (with value-differing variables indicated) observed at the divergence point in both buggy and patched executions.
We generate $N_{\text{candidate}}$ expressions for both changed and unchanged expressions (our experiments use $N_{\text{candidate}}=10$).

Given $N_{\text{candidate}}$ generated expressions, we validate each to ensure it exhibits the expected behavioral property. Specifically, a \textit{changed} expression must evaluate to different values in the buggy and patched programs, whereas an \textit{unchanged} expression must evaluate identically. We implement a debugger to evaluate each candidate in both programs and retain only those consistent with their assigned category, resulting in $N_{\text{changed}}$ and $N_{\text{unchanged}}$ validated candidates for the changed and unchanged categories, respectively.

\subsubsection{Candidate Expression Selection}
After obtaining validated candidate expressions, we finalize the multiple-choice options. We adopt the following strategy to avoid trivial/obvious choices. Let $C = \{c_1, \dots, c_{N_{\text{changed}}}\}$ denote the set of changed expressions and $U = \{u_1, \dots, u_{N_{\text{unchanged}}}\}$ the set of unchanged expressions.

We choose the correct option among the candidates $c_i \in C$ according to its average similarity to the unchanged candidates $u_i \in U$, as defined in Equation~\ref{eq:sim}.
\begin{gather}
\label{eq:sim}
\small
    \mathrm{sim}_{\text{avg}}(c, U) = \frac{1}{|U|} \sum_{u \in U} \mathrm{sim}(c, u)  
\end{gather}
We use the normalized Levenshtein distance~\cite{yujian2007normalized} for the similarity $\mathrm{sim}(c, u)$. %
This favors candidates $c_{\text{selected}} \in C$ with high similarity to unchanged expressions, which are thus easily confusable with unchanged ones, reducing the likelihood of trivial questions. 

To select distractors, we leverage a maximal marginal relevance (MMR) criterion~\cite{carbonell1998use}. The goal is to discourage selecting multiple distractors that are near-duplicates of each other while still encouraging distractors similar to $c_{\text{selected}}$. Each unchanged candidate $u_i \in U$ is scored as below:
\begin{equation}
\label{eq:mmr}
\small
    \mathrm{MMR}(u) = \lambda \cdot \mathrm{sim}(u, c_{\text{selected}})
\;-\;
(1-\lambda) \cdot \max_{d \in D} \mathrm{sim}(u, d)
\end{equation}
Here, $D$ denotes the set of already selected distractors and $\lambda \in [0,1]$ controls the relevance-diversity trade-off ($\lambda$ defaults to 0.7 in our experiments). \ys{The similarity for both $\mathrm{sim}(u, c_{\text{selected}})$ and $\mathrm{sim}(u, d)$ is computed using Levenshtein distance}. The top $N_\text{distractor}$ expressions are selected iteratively. 

\subsection{Question Quality Assurance}

To ensure the quality of our generated questions, we perform the following quality checks. For the end-to-end questions, we ensure that PBTs fail on the buggy version and pass upon applying the developer patch. For the local questions involving expressions, each delta behavior was manually checked to verify that it reflects a genuine behavioral change from the patch. Two authors jointly reviewed and discussed all uncertain instances to reach consensus. 
For all questions, we check that instructions are clear by running \name on blank explanations and ensuring that this yields a score of 0 (as the LLMs pick the incorrect option ``\textit{explanation insufficient to answer}''). As an additional quality check, we performed a manual examination through an assessment procedure based on prior work~\cite{yang2022predictive}. Specifically, two authors independently answered 40 randomly sampled questions covering all question types from \name. Any disagreements were resolved by a third author. Agreement between human and LLM answers reached a Cohen's Kappa~\cite{cohen1960coefficient} of 0.7, which is considered as substantial agreement~\cite{le2019reliability, landis1977measurement, risse2025top}. We show an example of human labeling and disagreement resolution in Figure~\ref{fig:qa_disagreement_example}.

\begin{figure}
    \centering
    \begin{minted}[escapeinside=||,fontsize=\scriptsize,breaklines,breakanywhere]{text}
|\textbf{Explanation:}|
All the relevant tests pass. The issue was that passing a uint8 array to a colormap makes Matplotlib assign values like 256, 257, 258. These are out of range for uint8, causing NumPy 1.24+ deprecation warnings. The fix casts xa to int. Testing showed no more warnings. ...
|\textbf{Test Content:}|
...
|\textbf{Question:}|
Q: What happens before and after applying the patch?
 ...
 C. Fail at line 22 with TypeError.
 D. Fail at line 27 with AssertionError.
 E. Test passed.
|\textbf{Labeler 1: C,E}|
Line 22 fails because Matplotlib tries to assign 256--258 into a uint8 array. After the patch, xa is cast to int, so the test passes.
|\textbf{Labeler 2: D,E}|
The explanation says the bug emits DeprecationWarnings. Before the patch, dep_warnings should be non-empty, so line 27 fails. After the patch, the test passes.
|\textbf{Adjudicator: D,E}|
The explanation identifies warnings as the key buggy behavior and says the patch yields no more warnings. It does not justify inferring a TypeError at line 22. Thus, line 27 fails before the patch, and the test passes after the patch.
    \end{minted}
    \vspace{-0.3cm}
    \caption{Example of a resolved human-labeler disagreement for the end-to-end question.}
    \label{fig:qa_disagreement_example}
    \vspace{-0.3cm}
\end{figure}

\section{Auditing Code Explanations from Agents}
\label{sec:audit_agent}

Beyond evaluating explanation quality with \name, we further seek a general approach to improve explanations produced by coding agents, to better establish trust between agents and developers in future software development processes. To this end, we propose \auditagentname, which audits explanations from arbitrary coding agents and provides a refined explanation.

\noindent\textbf{\textit{Agent Operation.}}
\ys{Given an agent patch, agent explanation, and original issue description as input, \auditagentname first generates tests to double-check the explanation. For each generated test, \auditagentname invokes its \textsc{DiffExecution} tool, which executes the generated test before and after the patch. \textsc{DiffExecution} enables the agent to inspect the patch effect by reporting the test results, such as process return code or \texttt{stdout} content. \auditagentname can also invoke its \textsc{InspectCallGraph} tool to examine the call chain leading to the execution of a target function. Moreover, \auditagentname has access to other tools: a bash interface, file I/O utilities, and patch application/reversion to support the testing. After conducting differential testing and call graph inspection, \auditagentname compares the input agent explanation against the collected evidence from the invoked tools. If the evidence contradicts the explanation, \auditagentname revises the explanation, explicitly noting the conflicting evidence and how it contradicts the original explanation. On the other hand, if the evidence aligns with the explanation, \auditagentname appends a confirmation detailing the validation steps performed and explaining why the evidence supports the claim. Overall, \auditagentname augments agent explanations with independent, execution-based evidence from testing that the issue-resolving agent may have overlooked.}

\section{Experimental Setup}
\label{sec:expr_setup}

\subsection{Research Questions}

\hspace{\parindent}\textbf{RQ1: How do different agents score on \name?} Our work automatically evaluates the explanation quality of agents. As such, we compare how explanations from different agents score on \name, and how it differs from agent efficacy (i.e., the number of resolving patches generated).

\textbf{RQ2: What are the current problems with agent explanations?} Using \name, we analyze what specific problems the current agent explanations have: are low scores on \name due to vagueness or are explanations confidently wrong?

\textbf{RQ3: How does \auditagentname improve explanations for different agents?} We evaluate to what degree \auditagentname can enhance explanations from the issue-resolving agents we evaluate, as a showcase of how explanations can be made more trustworthy.

\subsection{Evaluated Agents}

\begin{table}[t]
  \caption{Evaluated agent frameworks.}
  \label{tab:agent-list}
  \centering
  \small
  \begin{tabular}{@{}lrccc@{}}
    \toprule
    \multicolumn{1}{l}{\textbf{\faGithub~Project}} &
    \multicolumn{1}{c}{\textbf{Stars}} &
    \multicolumn{1}{c}{\textbf{Paper}} &
    \multicolumn{1}{c}{\textbf{Commercial}} \\
    \midrule
    OpenHands/OpenHands~\cite{wang2024openhands} 
    & 67.3K & \checkmark & \checkmark \\
    bytedance/trae-agent~\cite{traeresearchteam2025traeagentllmbasedagent} 
    & 10.7K & \checkmark & \checkmark \\
    lingxi-agent/Lingxi~\cite{yang2025lingxirepositorylevelissueresolution}
    & 208   & \checkmark & --         \\
    smallcloudai/refact~\footnote{\url{https://github.com/smallcloudai/refact}}
    & 3.5K  & --         & \checkmark \\
    SWE-agent/mini-SWE-agent~\footnote{\url{https://github.com/SWE-agent/mini-swe-agent}}  
    & 2.7K  & --         & --         \\
    \bottomrule
  \end{tabular}
  \vspace{-0.3cm}
\end{table}

We evaluate a set of representative open-scaffold agent frameworks on our benchmark, as shown in Table~\ref{tab:agent-list}. Candidate agents are first identified from the top-20 submissions on the SWE-bench Verified leaderboard\footnote{\url{https://www.swebench.com}} as of February 2026. From this pool, we apply the following selection criteria.
First, the agent must either demonstrate substantial community adoption, as reflected by its GitHub star count, or be accompanied by a publicly available paper or technical report.
Second, complete issue-resolving trajectories for the agent must be publicly available.
Third, for at least 90\% of bug instances, the agent’s execution trajectory must include a patch explanation in the final tool invocation step.
In the end, we included the five open-scaffold agents in Table~\ref{tab:agent-list}.

\subsection{Implementation Details}

\begin{figure}[h]%
    \centering
    \vspace{-0.2cm}
    \begin{minted}{markdown}
An AI agent fixed a bug in a code repository and provided an explanation for the patch. You will be given this patch explanation, and your task is to answer questions about the bug and patch described by the explanation. ...

|\textbf{Patch Explanation:}|
|\textcolor{gray}{*** Explanation Start ***}|
|\textbf{\textcolor{blue}{[explanation]}}|
|\textcolor{gray}{*** Explanation End ***}|

|\textbf{\textcolor{blue}{[context]}}|

|\textbf{Question:}|
|\textbf{\textcolor{blue}{[question]}}|
    \end{minted}
    \vspace{-0.2cm}
    \caption{Question-answering prompt template.}
    \label{fig:prompt_template}
    \vspace{-0.2cm}
\end{figure}

\subsubsection{Prompt Design}
We evaluate patch explanations by prompting an LLM in a question-answering format. Fig.~\ref{fig:prompt_template} illustrates the prompt template in \name, which is applied across all question types. For each specific question type, we populate the template with an explanation, the corresponding context, and the multiple-choice question, as introduced in Section~\ref{sec:methodology:formats}. To populate the \texttt{[context]} placeholder, we use the generated PBTs for end-to-end questions; for local questions, the pre-patch function containing delta behavior, its inputs, and the corresponding line of divergence.
For the \texttt{[question]} placeholder, Fig.~\ref{fig:example_qs} presents an example for each question type.

\begin{table}[t]
\caption{SWE-bench Verified Issue Exclusion Categories}
\label{tab:benchmark_statistics}
\small
\begin{tabularx}{0.45\textwidth}{@{}Xr@{}}
\toprule
\textbf{Reason}                                                     & \textbf{\# Excluded} \\ \midrule
SWE-bench harness failing with developer patches                    & 13                             \\
Complexity of traced program execution                              &                                \\
\quad - Large traces (can be up to ~70 GB for one test case)        & 6                              \\
\quad - 100\% CPU usage                                             & 1                              \\
\quad - Timeout (after 6 hours)                                     & 14                             \\
Idiosyncratic tests in SWE-bench Verified                           &                                \\
\quad - Buggy program failing before test function is called        & 14                             \\
\quad - Hard-to-handle dynamic features in Python                   & 12                             \\
\quad - Repository-specific custom test framework features          & 7                              \\
Fragile program states (logging causes test failure)                & 67                             \\
Question quality control                                            &                                \\
\quad - Delta behavior in variables with random values              & 16                             \\
\quad - Delta behavior in fields in deeply nested objects           & 30                             \\
\quad - Expression generation failure after multiple trials         & 23                             \\ \midrule
\textbf{Total}                                                      & 203                            \\ \bottomrule
\end{tabularx}
\vspace{-0.2cm}
\end{table}

\subsubsection{Benchmark Statistics}

In this work, we reuse bug instances from SWE-bench Verified~\cite{chowdhury2024swebenchverified}, a widely-adopted benchmark of 500 real-world software engineering issues. Although SWE-bench Verified is human-validated, a small number of bug instances fail to pass the test suites even when developer patches are applied, as reported by 
many including Anthropic\footnote{\url{https://www.anthropic.com/news/claude-3-7-sonnet}}. Therefore, we exclude these instances in our study.
To ensure reliable instrumentation, we further exclude instances whose execution exceeds practical tracing limits due to excessive traces, runtime, CPU utilization, or other problems.
Based on these factors, we construct \name based on 297 bug instances. The reasons for excluding the remaining bug instances are summarized in Table~\ref{tab:benchmark_statistics}. Specifically, ``fragile program states'' refer to implicit side effects during serialization-based logging. We acknowledge this and the tracing limits as limitations of the data construction pipeline. Nevertheless, we verify that the exclusion of bugs does not lead to statistically significant differences between \name and SWE-bench in terms of project composition, human-rated difficulty, and patch size.

\subsubsection{Configurations and Environment}

To obtain patch explanations and efficacy scores for the agents, we collect agent trajectories and test execution logs from \url{ https://github.com/SWE-bench/experiments}. 
GPT-5.2~\cite{gpt52} was used to generate bug-reproducing PBTs and candidate expressions;
GPT-5-mini~\cite{gpt5mini} is used as the question-answering LLM. We use the weaker GPT-5-mini model to answer the questions as we want the model to rely on the explanation, whereas better models may use their superior knowledge to answer the questions. 
\auditagentname also uses GPT-5-mini. 
All models are accessed via OpenAI API
with default options: thus, the question answering LLM, GPT-5-mini, uses a temperature of 1.0 and medium reasoning effort.
For each question, we generate the answer five times and report the average explanation score across these generations.

\begin{table*}[t]
\centering
\caption{Evaluation results of agents on \name, compared with efficacy results on SWE-bench Verified. For explanation quality, we report scores on four individual evaluation components, and the final explanation scores with standard error of the mean (SEM). The same subset used in \name evaluation is used to evaluate efficacy. The numbers after the \# character indicate the agent ranking in the respective column.}
\label{tab:eval_results}
\small
\begin{tabular}{@{}crrrrcc@{}}
\toprule
\multirow{3}{*}{\textbf{Agent}} &
  \multicolumn{5}{c}{\textbf{\name}} &
  \multicolumn{1}{c}{\textbf{SWE-bench Verified}} \\ \cmidrule(l){2-7}
 &
  \multicolumn{2}{c}{\textbf{End-to-End}} &
  \multicolumn{2}{c}{\textbf{Local}} &
  \multicolumn{1}{c}{\multirow{2}{*}{\textbf{Expl. Score}}} &
  \multicolumn{1}{c}{\multirow{2}{*}{\textbf{Efficacy}}} \\ \cmidrule(lr){2-5}
 &
  \multicolumn{1}{c}{\textit{Intent}} &
  \multicolumn{1}{c}{\textit{Effect}} &
  \multicolumn{1}{c}{\textit{Intent}} &
  \multicolumn{1}{c}{\textit{Effect}} &
  \multicolumn{1}{c}{} &
  \multicolumn{1}{c}{} \\ \midrule
refact         & 0.723 & 0.818 & 0.355 & 0.490 & 0.596 {\footnotesize $\pm 0.006$} \textcolor{gray}{(\#2)} & 0.778 \textcolor{gray}{(\#3)} \\
Lingxi         & 0.704 & 0.805 & 0.368 & 0.492 & 0.592 {\footnotesize $\pm 0.005$} \textcolor{gray}{(\#3)} & 0.785 \textcolor{gray}{(\#2)} \\
OpenHands      & 0.723 & 0.791 & 0.353 & 0.522 & 0.597 {\footnotesize $\pm 0.003$} \textcolor{gray}{(\#1)} & 0.727 \textcolor{gray}{(\#4)} \\
trae-agent     & 0.636 & 0.805 & 0.332 & 0.457 & 0.558 {\footnotesize $\pm 0.005$} \textcolor{gray}{(\#4)} & 0.818 \textcolor{gray}{(\#1)} \\
mini-SWE-agent & 0.467 & 0.599 & 0.302 & 0.370 & 0.435 {\footnotesize $\pm 0.006$} \textcolor{gray}{(\#5)} & 0.599 \textcolor{gray}{(\#5)} \\ \bottomrule
\end{tabular}
\vspace{-0.2cm}
\end{table*}

\section{Results}
\label{sec:results}

\subsection{RQ1: Agent Explanation Scores}
\label{sec:results_rq1}

\noindent\textbullet\quad\textbf{Patch explanation quality is distinct from patch generation efficacy.} Our evaluation of the explanations of five agents on \name, along with the efficacy scores on SWE-bench, is presented in Table~\ref{tab:eval_results}.
Note that higher efficacy on SWE-bench does not necessarily correspond to better explanation quality. OpenHands, which ranks fourth in efficacy, achieves the highest explanation score among all evaluated agents. In contrast, trae-agent, which attains the highest efficacy, ranks only fourth in explanation quality. 
These shifts in ranking highlight the importance of independently evaluating explanation quality.

\noindent\textbullet\quad\textbf{Results from \name are stable.} Although our benchmark involves an LLM in the evaluation process, the results from the benchmark are stable, with a standard error of the mean of less than 0.01 for the explanation scores in Table~\ref{tab:eval_results}. Furthermore, running \name with a different question-answering LLM (GPT-5-nano) yields an identical explanation score ranking; results for this experiment can be found in our supplementary material.

\noindent\textbullet\quad\textbf{Explanations describe end-to-end behavior better than local behavior.} We find that for both intent and effect, end-to-end scores are consistently higher than local scores across all agents. This indicates that current agent explanations are more informative at global rationales than localized code-level reasoning.

\vspace{-0.2cm}
\subsection{RQ2: Agent Explanation Problem Analysis}
\label{sec:results_rq2}

\begin{table*}[t]
  \centering
  \caption{Failure breakdown of agent-generated explanations. Percentages denote the fraction labeled as \textit{misaligned} (incorrect/misleading) or \textit{not-informative} (insufficient information).}
  \label{tab:failure_modes_combined}
  \setlength{\tabcolsep}{5pt}
  \small
  \begin{tabular}{c c c c c c c c c}
    \toprule
    \multicolumn{1}{c}{} &
    \multicolumn{4}{c}{\textbf{End-to-End}} &
    \multicolumn{4}{c}{\textbf{Local}} \\
    \cmidrule(lr){2-5}\cmidrule(lr){6-9}
    \textbf{Agent} &
    \multicolumn{2}{c}{\textbf{Intent}} &
    \multicolumn{2}{c}{\textbf{Effect}} &
    \multicolumn{2}{c}{\textbf{Intent}} &
    \multicolumn{2}{c}{\textbf{Effect}} \\
    \cmidrule(lr){2-3}\cmidrule(lr){4-5}\cmidrule(lr){6-7}\cmidrule(lr){8-9}
    &
    \textit{Mis.} & \textit{NI} &
    \textit{Mis.} & \textit{NI} &
    \textit{Mis.} & \textit{NI} &
    \textit{Mis.} & \textit{NI} \\
    \midrule
    refact         & 4.0\% & 23.7\% & 13.7\% & 4.5\%  & 37.4\% & 27.1\% & 39.8\% & 11.2\% \\
    Lingxi         & 4.6\% & 25.0\% & 14.9\% & 4.5\%  & 36.2\% & 26.9\% & 41.5\% & 9.3\%  \\
    OpenHands      & 3.2\% & 24.5\% & 17.9\% & 3.0\%  & 35.5\% & 29.2\% & 40.0\% & 7.8\%  \\
    trae-agent     & 4.6\% & 31.8\% & 14.7\% & 4.7\%  & 37.3\% & 29.5\% & 39.4\% & 14.9\% \\
    mini-SWE-agent & 3.4\% & 49.8\% & 25.4\% & 14.7\% & 29.9\% & 39.9\% & 38.9\% & 24.1\% \\
    \bottomrule
  \end{tabular}
\end{table*}

\begin{table}[t]
  \centering
  \caption{Overconfidence in end-to-end effect predictions. Overconfidence is the percentage of non-passing patches for which the Q\&A LLM incorrectly predicts that the bug-reproducing test would pass after applying the patch.}
  \label{tab:overconfidence_effect}
  \small
  \begin{tabular}{lrr}
    \toprule
    \textbf{Agent} & \textbf{\# Not Pass} & \textbf{Overconfidence} \\
    \midrule
    refact         & 66  & 78.79\% \\
    Lingxi         & 64  & 82.81\% \\
    OpenHands      & 81  & 71.60\% \\
    trae-agent     & 54  & 79.63\% \\
    mini-SWE-agent & 104 & 83.65\% \\
    \midrule
    \textbf{Average} & \textbf{73.8} & \textbf{79.30\%} \\
    \bottomrule
  \end{tabular}
\end{table}

We analyze explanation quality further through two failure modes: \textit{uninformative} explanations and \textit{misaligned} explanations. Uninformative explanations are cases where the LLM-selected answer is ``\textit{Explanation insufficient to answer}'' (provided at the end of each question), indicating a lack of information.
Misalignment refers to cases where the LLM-selected answer differs from the ground-truth answer and is not classified as uninformative, indicating that the explanation leads to an incorrect conclusion.

\noindent\textbullet\quad\textbf{End-to-end intent in explanations is generally accurate but sometimes absent from agent explanations.} Table~\ref{tab:failure_modes_combined} shows that failures on the end-to-end intent explanations are dominated by non-informativeness, with misalignment remaining uniformly low. That is, when agent explanations describe what the program as a whole should do, it is generally accurate. However, they omit intent at times, in favor of reporting what the agent patch does.

\noindent\textbullet\quad\textbf{Explanations frequently inaccurately describe local intent.} For local intent, misalignment increases substantially relative to end-to-end intent. Local intent requires precisely describing what behavior a function should show. Even when end-to-end intent is correctly described, agents appear to inaccurately infer what the developer intended at the local level.

\noindent\textbullet\quad\textbf{Explanations are overconfident in the correctness of their patch.} For end-to-end effect, misalignment is the dominant failure case across all agents.
We find that agent explanations almost always describe what the patch does, but frequently do so incorrectly. In Table~\ref{tab:overconfidence_effect}, across all agents, 79.30\% of incorrect patches are nevertheless predicted by the LLM to \emph{pass} the bug-reproducing test in the end-to-end effect question.

\begin{figure*}
    \centering
    \begin{subfigure}[b]{0.48\textwidth}
        \centering
        \begin{minipage}{\textwidth}

            \begin{subfigure}[t]{\textwidth}
                \centering
                \begin{minted}[escapeinside=$$,fontsize=\scriptsize,breaklines,breakanywhere]{diff}
- if app_dir in fixture_dirs:
+ if app_dir in [str(d) for d in fixture_dirs]:
    raise ... 
                \end{minted}
                \vspace{-0.2cm}
                \caption{The intended change.}
                \label{fig:dev_patch_rq2}
            \end{subfigure}
            \begin{subfigure}[t]{\textwidth}
                \centering
                \begin{minted}[escapeinside=$$,fontsize=\scriptsize,breaklines,breakanywhere]{diff}
- if len(fixture_dirs) != len(set(fixture_dirs)):
+ normalized_dirs = []
+ for d in fixture_dirs:
+   try:
+       normalized = str(os.path.realpath(...)
+       normalized_dirs.append(normalized)
+   except (OSError, ValueError):
+       ...
+ if len(normalized_dirs) != len(set(normalized_dirs)):
    ...
  if app_dir in fixture_dirs:
    raise ...
        \end{minted}
        \vspace{-0.2cm}
                \caption{The agent patch.}
                \label{fig:agent_patch_rq2}
            \end{subfigure}

        \end{minipage}
        \label{fig:intended_change}
    \end{subfigure}%
    \begin{subfigure}[b]{0.48\textwidth}
        \centering
        \begin{minted}[fontsize=\scriptsize,breaklines]{markdown}
### **Changes Applied:**
...
**Modified** `.../commands/loaddata.py` - Fixed the duplicate detection logic to properly handle Path instances
...
### **Comprehensive Test Results:**
1. **Fixtures Regress Tests**: **59/59 tests passed** (including 2 new tests)
...
### **Functionality Verified:**
...
- **Mixed Path/string duplicates** are now properly detected  
- **Valid Path usage** continues to work without false positives
...
The GitHub issue **"Fixture dirs duplicates undetected if dir is Path instance"** has been **completely resolved** with a robust, backward-compatible solution that properly normalizes paths before duplicate detection.
        \end{minted}
        \caption{Agent explanation.}
        \label{fig:agent_explanation}
    \end{subfigure}
    \vspace{-0.3cm}
    \caption{Example of explanation misalignment.}
    \label{fig:rq2_case_study}
    \vspace{-0.2cm}
\end{figure*}

\subsubsection{Case Study}
Fig.~\ref{fig:rq2_case_study} presents a case study of a low-quality agent-generated explanation and its assessment by \name.
To summarize, the intended change is to ensure a function accounts for type heterogeneity by converting all elements of `\texttt{fixture\_dirs}' to strings before comparing with `\texttt{app\_dir}', as shown in Fig.~\ref{fig:dev_patch_rq2}. When the bug is fixed, an \texttt{ImproperlyConfigured} exception should be triggered when \texttt{app\_dir} is included in \texttt{fixture\_dirs}.
Instead, the agent patch changes the irrelevant logic of duplicate entry detection in the function, and thus does not fix the bug.%

Our main focus, the explanation in Fig.~\ref{fig:agent_explanation}, is also problematic, which \name automatically reveals. For example, the explanation is uninformative in \emph{end-to-end intent}: it only notes that duplicate detection logic is now `properly handled', without referencing any specifics of the intended behavior change. 
Accordingly, when \name evaluates this explanation with respect to end-to-end intent, the LLM-selected answer is that the explanation is insufficient to complete the PBT. 
Meanwhile, the end-to-end and local effect descriptions in the explanation are \emph{misaligned} as the explanation falsely claims to have fixed the issue.%
When \name evaluates the explanation for the end-to-end and local effect questions, the LLM-selected answers are that the patch makes the test pass and that the patch leads to a change in return values; both are wrong as the patch has no effect. Thus, the explanation is marked as misaligned in effect by \name.

\subsection{RQ3: Improving Explanations}
\label{sec:results_rq3}

\begin{table*}[t]
    \centering
    \caption{Improvement of explanations on \name score after \auditagentname auditing. The numbers in parentheses represent the score gain relative to non-audited explanations as shown in Table~\ref{tab:eval_results}.}
\small
\scalebox{1.0}{
\begin{tabular}{@{}clllll@{}}
\toprule
\multirow{2}{*}{\textbf{Audited Agent}} &
  \multicolumn{2}{c}{\textbf{End-to-End}} &
  \multicolumn{2}{c}{\textbf{Local}} &
  \multicolumn{1}{c}{\multirow{2}{*}{\textbf{Expl. Score}}} \\ \cmidrule(lr){2-5}
 &
  \multicolumn{1}{c}{\textit{Intent}} &
  \multicolumn{1}{c}{\textit{Effect}} &
  \multicolumn{1}{c}{\textit{Intent}} &
  \multicolumn{1}{c}{\textit{Effect}} &
  \multicolumn{1}{c}{} \\ \midrule
refact         & 0.768 (+6.2\%) & 0.837 (+2.3\%) & 0.359 (+1.1\%) & 0.504 (+2.9\%) & 0.617 (+3.4\%) \\
Lingxi         & 0.768 (+9.1\%) & 0.813 (+1\%) & 0.362 (-1.6\%) & 0.501 (+1.8\%) & 0.611 (+3.2\%) \\
OpenHands      & 0.774 (+7.1\%) & 0.806 (+1.9\%) & 0.366 (+3.7\%) & 0.516 (-1.1\%) & 0.616 (+3.1\%) \\
trae-agent     & 0.782 (+23.0\%) & 0.826 (+2.6\%) & 0.361 (+8.7\%) & 0.512 (+12\%) & 0.620 (+11.3\%) \\
mini-SWE-agent & 0.718 (+56.1\%) & 0.731 (+22.0\%) & 0.362 (+19.9\%) & 0.510 (+37.8\%) & 0.580 (+33.5\%) \\ \bottomrule
\end{tabular}
}
    \label{tab:rq3_improval}
    \vspace{-0.2cm}
\end{table*}

We apply \auditagentname to improve patch explanations for all agents evaluated in RQ1, at a cost of \$0.05 per explanation on average. The results of running \auditagentname are shown in Table~\ref{tab:rq3_improval}. The auditing process improved the explanation score for all agents evaluated, with notable improvement for end-to-end questions. This is expected, as \auditagentname primarily modifies explanations by executing tests, which are an end-to-end interaction with the program. While the expectation when developing \auditagentname was that the effect score would improve, there was also notable improvement for the end-to-end intent scores of all agents. We find that as \auditagentname runs tests, it reasons about the expected behavior of the program and includes it in its final explanation. Thus, it addresses the uninformative problem for end-to-end intent questions reported in Table~\ref{tab:failure_modes_combined}.  %

\ys{As an example, \auditagentname improved the explanation in the case study of RQ2. As noted before, the explanation claims the bug is fixed when it is not. \auditagentname identifies this discrepancy through testing, then adds to the original explanation the following:}
\vspace{-0.15cm}
\begingroup
\addtolength\leftmargini{-0.2in}
\begin{quote}
    \ys{\textit{(important caveat): ... If FIXTURE\_DIRS contains a pathlib.Path object equal to app\_dir, the patch will not detect it and will not raise the ImproperlyConfigured error. I reproduced this behavior: (...)
    }}
\end{quote}
\vspace{-0.15cm}
\endgroup
\noindent 
\ys{This refined explanation accurately describes the patch effect and intended behavior. As such, the end-to-end questions of \name are now accurately answered, improving explanation score.}

To better interpret the gains from \auditagentname, we sampled 60 explanations from the two agents most affected by \auditagentname, namely mini-SWE-agent and trae-agent, and manually categorized the changes introduced by \auditagentname. The results show that \auditagentname generally added information about ``\textit{how the code behaves before and after the patch}'' (46\%), ``\textit{which parts of the code remained unchanged}'' (26\%), and ``\textit{edge cases}'' (5\%). Also, in 17\% of the cases, \auditagentname added example values. These findings suggest that \auditagentname improves not only performance under the test-oriented scoring format but also the usefulness of explanations for developer comprehension.

\section{Implications}

Based on our empirical findings, we make the following recommendations to agent developers interested in improving the quality of their agent's explanations.

\noindent\textbullet\quad\textbf{Structurally encourage explanations.} A key finding from \name is that agents such as OpenHands can generate better explanations despite having lower patch resolution rates. Upon analyzing OpenHands, we find that OpenHands architecturally enforces explanations through its \texttt{finish} tool, which must be run to submit a solution. The tool description states that \texttt{finish} should be provided with a \texttt{message} parameter which includes ``a clear summary of actions taken and their results''. In contrast, trae-agent, which has a generally similar architecture to OpenHands, critically does not include a \texttt{message} parameter in its \texttt{finish} tool. This leads to explanations being short or completely absent.

\noindent\textbullet\quad\textbf{Clearly describe explanation desiderata in the system prompt.} The system prompt also plays an important role in determining the quality of explanations. The agent with the second-highest score on \name, refact, directs the agent to ``post a concise, bullet-point summary that includes the suspected root cause''. In contrast, mini-SWE-agent does not mention the need of explanations anywhere in its prompt. Along with the fact that mini-SWE-agent does not structurally encourage explanations, this leads to mini-SWE-agent having the worst explanation score among evaluated agents.

\noindent\textbullet\quad\textbf{Employ an independent explanation auditing sub-agent.} As RQ3 shows, the explanations of all agents improved in quality when validated by \auditagentname. Thus, especially for multi-agent systems such as Lingxi, it is worth considering including a sub-agent dedicated to auditing the natural language explanation of a coding agent.

\section{Threats to Validity}
\label{sec:discussion}

\noindent\textbf{\textit{External Validity.} }
The benchmark and findings of this work are limited to a set of open-source projects from SWE-bench Verified, which contains repositories with regression test suites and reproducible failures. Nonetheless, the benchmark framework is not inherently limited to SWE-bench and can be extended easily.

\noindent\textbf{\textit{Internal Validity.}}
Explanation scores are influenced by stochastic variation. To minimize this threat, we run each experiment five times and report aggregated results. As shown in Table~\ref{tab:eval_results}, the standard error of the mean is low, indicating that the results are stable. 

\noindent\textbf{\textit{Construct Validity.}}
Explanation score captures only a subset of dimensions relevant to explanation quality, such as behavioral alignment, and does not assess qualitative aspects including readability or usefulness. While acknowledging these limitations, we adopt this design choice to enable easily scalable and reproducible evaluation.

\section{Conclusion}
\label{sec:conclusion}
As LLM agents play a greater role in software development, the explanations they generate will play an increasingly critical role in determining developer trust. Despite this,
a systematic evaluation of their quality has not been done. In this work, we present \name, a benchmark to automatically evaluate agent explanations. \name is based on the intuition that informative explanations would help correctly answer questions about the bug and agent patch. Evaluation on \name revealed that agent explanation quality and patch generation efficacy do not always correspond, affirming the need to evaluate explanations. Furthermore, \name revealed the specific ways in which agent explanations could be improved, such as overconfidence in patch correctness. With this in mind, we implement \auditagentname to autonomously run tests and calibrate overconfident explanations. \auditagentname improved agent scores on \name by 10.9\% on average, suggesting a mechanism through which explanation quality can be automatically improved.

\begin{acks}
This research is supported by the National Research Foundation, Singapore, under NRF Investigatorship Program, Award ID: NRF-NRFI11-2026-0001, ``Agentic AI based Software of the future: from Scale to Trust''.
\end{acks}

\section*{Data Availability Statement}

Our replication package is available at~\cite{explainbench_package}. We provide a leaderboard website of \name at \url{https://explainbench.github.io}.

\onecolumn
\begin{multicols}{2}
\bibliographystyle{ACM-Reference-Format}
\bibliography{ref}
\end{multicols}

\end{document}